\documentstyle[aps,epsfig,preprint]{revtex}

\begin{document}

\title{Coherent carrier dynamics in semiconductor superlattices}

\author{Enrique Diez,$^{a}$ Rafael G\'omez-Alcal\'a,$^{b}$
Francisco\ Dom\'{\i}nguez-Adame,$^{c}$ Angel S\'{a}nchez,$^{a}$
and Gennady P.\ Berman$^{d}$}

\address{$^a$GISC, Departamento de Matem\'aticas, Universidad Carlos
III, E-28911 Legan\'{e}s, Madrid, Spain\\
$^b$Departamento de Tecnolog\'{\i}as de las Comunicaciones, 
Universidade de Vigo, E-36200 Vigo, Spain \\
$^c$GISC, Departamento de F\'{\i}sica de Materiales,
Universidad Complutense, E-28040 Madrid, Spain\\
$^d$Theoretical Division and CNLS, B213, Los Alamos National
Laboratory, Los Alamos, New Mexico 8754 and Kirensky Institute of Physics, 660036 Krasnoyarsk, Russia }

\date{\today}

\maketitle

\begin{abstract}

We investigate the coherent dynamics of carriers in semiconductor
superlattices driven by ac-dc electric fields.  We solve numerically the
time-dependent effective-mass equation for the envelope function.  We
find that carriers undergo Rabi oscillations when the driving frequency
is close to the separation between minibands.

\end{abstract}


\narrowtext

Recent advances in laser technology make possible to drive semiconductor
nanostructures with intense coherent ac-dc fields.  This opens new
research fields in time-dependent transport in mesoscopic
systems~\cite{Pieper,Cundiff} and puts forward the basis for a new
generation of ultra-high speed devices.  Artificial two-level
semiconductor nanostructures acting as switching devices based on
Rabi oscillations (RO's) have been already suggested~\cite{Martin}.
Moreover, resonant phonon-assisted tunneling through a double quantum
dot could be used as an efficient electron pump from spatial
RO's~\cite{Stafford}.

Zhao {\em et al.\/} have investigated analytically a tight-binding model
of a two-band system in a time-dependent ac-dc field in the weak
coupling limit~\cite{Zhao}.  They identified RO's between Bloch bands
under resonant conditions, which reveal the existence of quasienergy
bands and fractional Wannier-Stark ladders.  The advances achieved in
molecular beam epitaxy, which allow to fabricate semiconductor
superlattices (SL's) tailored with the desired conduction-band profiles,
make these systems ideal candidates to propose experiments on coherent
carrier dynamics.  However, the tight-binding approximation presents
some limitations to describe actual semiconductor superlattices (SL's)
when the coupling between neighboring quantum wells is not weak.  Thus,
in order to experimentally access the validity of theoretical
predictions, one should use a more realistic model.  In this Letter we
present an effective-mass model beyond the tight-binding approximation,
containing all ingredients of actual SL's, namely finite interband
coupling and multiband scattering.

We consider electron states close to the conduction-band edge and use
the effective-mass approximation.  The electron wave packet satisfies the
following equation
\begin{eqnarray}
& & i\hbar\,\frac{\partial\Psi(x,t)}{\partial t} = \nonumber \\
& & \left[ -\,{\hbar^2\over 2m^{*}}\,{d^2\phantom{x}\over dx^2} +
V_{\mathrm{SL}}(x) - eF\,x\,\sin(\omega_{\mathrm{ac}} t)  
\right]\,\Psi(x,t),
\label{1}
\end{eqnarray}
where $x$ is the coordinate in the growth direction, $F$ and $\omega$
are the strength and the frequency of the ac field.  The SL potential at
flat band is $V_{\mathrm{SL}}(x) = \Delta E_c$ if $x$ lies inside the 
barriers and zero otherwise, $\Delta E_c$ being the conduction-band 
offset.  We have considered a constant effective-mass $m^{*}$ for simplicity.

The band structure at flat band is computed by using a finite-element
method~\cite{Gomez}.  The eigenstate $j$ of the band $i$ with
eigenenergy $E_{i}^{(j)}$ is denoted as $\psi_{i}^{(j)}(x)$.  A good
choice for the initial wave packet is provided by using a linear
combination of the eigenstates belonging to the first miniband.  For the
sake of clarity we have selected as the initial wave packet $\Psi(x,0) =
\psi_{i}^{(j)}(x)$, although we have checked that this assumption can be
dropped without changing our conclusions.  The subsequent time evolution
of the wave packet $\Psi(x,t)$ is calculated numerically by means of an
implicit integration schema~\cite{Diez}.  In addition to $\Psi(x,t)$ we
also compute the probability of finding an electron, initially in the
state $\Psi(x,0)=\psi_{i}^{(j)}(x)$, in the state $\psi_{k}^{(j)}(x)$
\begin{equation}
P_{ik}^{(j)}(t)=\int_{-\infty}^{\infty}\>dx\,\Psi^{*}(x,t)\psi_{k}^{(j)}(x).
\label{probab}
\end{equation}

We present here the results for a SL with $10$ periods of $100\,$\AA\
GaAs and $50\,$\AA\ Ga$_{0.7}$Al$_{0.3}$As with band offset $\Delta
E_{c}=250\,$meV and $m^{*}=0.067m$, $m$ being the free electron mass.
We consider electric field $F=25\,$kV/cm as a typical value, although
similar results are observed for other strengths.  Figure~\ref{fig1}(a)
displays $P_{01}^{(5)}(t)$ for $F=25\,$kV/cm at the resonant frequency
$\omega_{\mathrm{ac}} = (E_{1}^{(5)} - E_{0}^{(5)})/\hbar = 150\,$THz.
Thus, we are monitoring the transitions between the central state
($j=5$) in the first miniband to the central state in the second
miniband as a function of time.  We observe the occurrence of very well
defined RO's with an amplitude close to $0.3$.  Summing up the
probabilities of the rest of states in the second miniband, the
probability of finding the electron in this band is very close to unity
($\sim 0.99$).  The frequency of the RO's, obtained performing the fast
Fourier transform (FFT) of $P_{01}^{(5)}(t)$, is $\omega_{\mathrm{Rabi}}
= 19.18\,$THz.  The probability $P_{01}^{(5)}(t)$ is dramatically
reduced when the ac driving field is out of resonance, as shown in
Fig.~\ref{fig1}(b) for $\omega_{\mathrm{ac}} = 100\,$THz.  The FFT of
those date reveals no specific features besides the peak at the driving
frequency $\omega_{\mathrm{ac}}$.

In a pure two-level system, a straightforward perturbation calculation
yields $\omega_{\mathrm{Rabi}}=|F_{01}|/\hbar$ at resonance, where
$F_{01}$ is the matrix element of the perturbation between the ground
state and the first excited state.  Thus, $\omega_{\mathrm{Rabi}}$ is
linear in the electric field in a pure two-level system.  Although the
SL is not a pure two-level system, we realize that this linear
dependence still holds, as we can see in Fig.~\ref{fig2}.

As an estimation of the leakage current that one could observe in
electron pumping devices based in RO's in SL's, we have studied the
integrated density in the right part of the SL, defined as
\begin{equation}
P_T(t)=\int_{x_r}^{\infty}|\Psi(x,t)|^{2}dx,
\label{p_t}
\end{equation}
where $x_r$ is the coordinate of the right edge of the SL.
Figure~\ref{fig3} shows the results at the resonant frequency
$\omega_{\mathrm{ac}} = 150\,$THz as well as out of resonance when
$\omega_{\mathrm{ac}}=200\,$THz.  Under resonant conditions the wave
function is emitted by {\em bursts\/} from the SL region every time a RO
has been completed.  On the contrary, the tunneling across the whole SL
is negligible when the ac driving frequency is not very close to the
resonant one.

In summary, we have studied the coherent carrier dynamic in
semiconductor SL's driven by an intense ac-dc field.  We found that the
electron can perform RO's under resonant conditions as in pure two-level
systems.  The Rabi frequency of the oscillations depends linearly on the
strength of the electric field.  We have shown that electrons are
emitted by {\em bursts\/} under resonant conditions, whereas the
tunneling probability in vanishingly small out of resonance.  Therefore,
we suggest that semiconductor SL's driven by an intense ac-dc field may
be used as an efficient electron pumping device in THz science.
Finally, we should mention that we have not considered the role of
imperfections or any other scattering mechanism that could result in a
reduction of the carrier coherence.  Our preliminary results show that
scattering by interface roughness does not strongly modify the above
picture.  Further work along these lines is currently in progress.

E.\ D.\ thanks to Antti-Pekka Jauho illuminating conversations and his
warm hospitality at Mikroelektronik Centret where this paper was
written in part.  Work at Legan\'es and Madrid is supported by CICYT
(Spain) under projects~MAT95-0325 and DGES~PB96-0119.  E.\ D.\
gratefully acknowledges partial support from Fundaci\'on Universidad
Carlos III de Madrid.  G.\ P.\ B.\ gratefully acknowledges partial
support by the Defense Advanced Research Projects Agency.

\begin{figure}
\caption{The probability of finding an electron, initially situated in
the state $\psi_{0}^{(5)}$, in the state $\psi_{1}^{(5)}$ as a function
of time for $F=25\,$kV/cm, when the ac field (a) is tuned to the
resonant frequency $\omega_{\mathrm{ac}} = 150\,$THz and (b) is out of
resonance $\omega_{\mathrm{ac}} = 100\,$THz.  Note the different
vertical scales.}
\label{fig1}
\end{figure}

\begin{figure}
\caption{Rabi frequency as a function of the dc field when the driving
frequency is tuned to the resonant frequency $\omega_{\mathrm{ac}} =
150\,$THz}
\label{fig2}
\end{figure}

\begin{figure}
\caption{$P_T(t)$ as function of time at at the resonant frequency
$\omega_{\mathrm{ac}} = 150\,$THz (solid line) and out of resonance
$\omega_{\mathrm{ac}}=200\,$THz (dashed line).}
\label{fig3}
\end{figure}
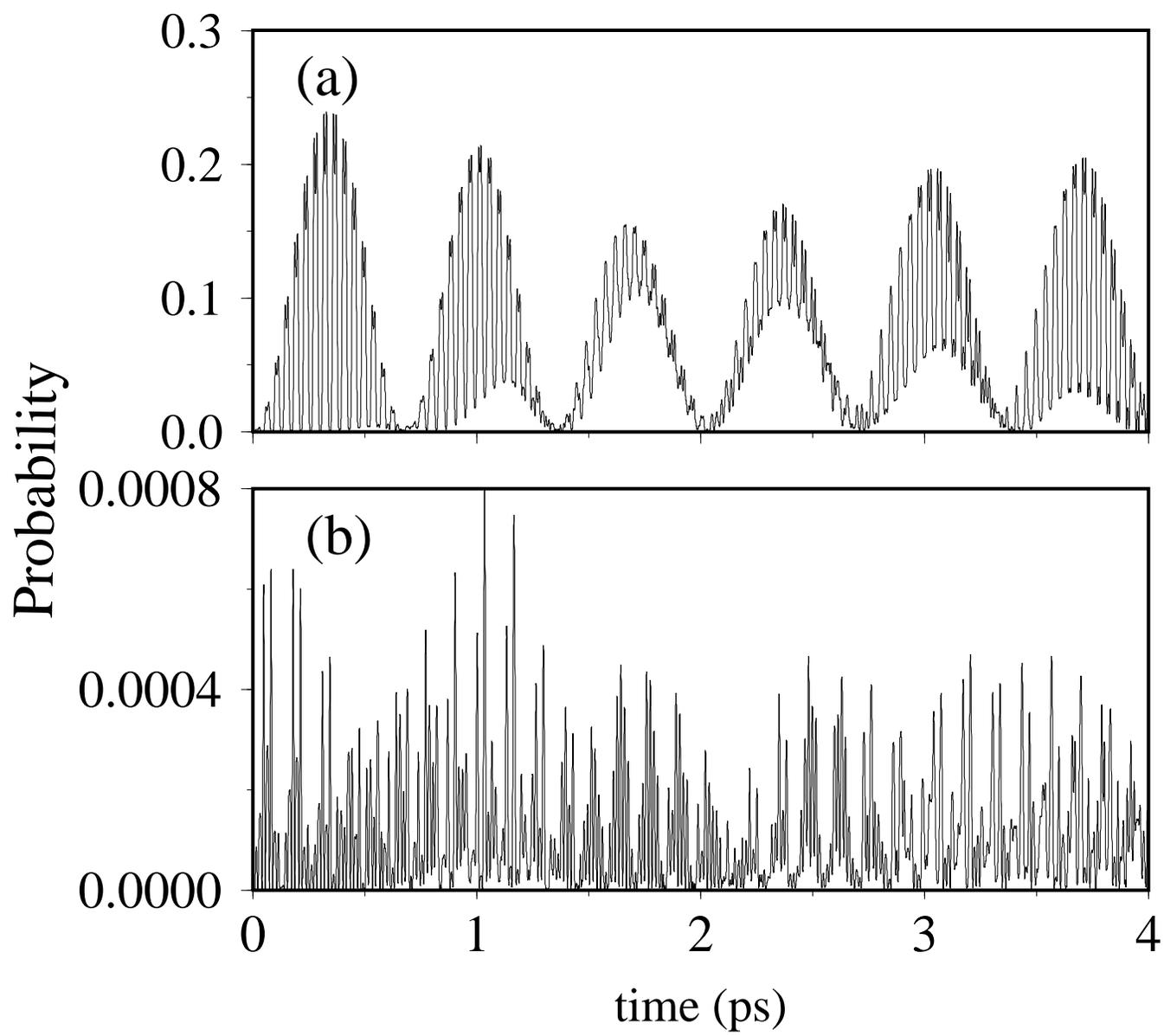
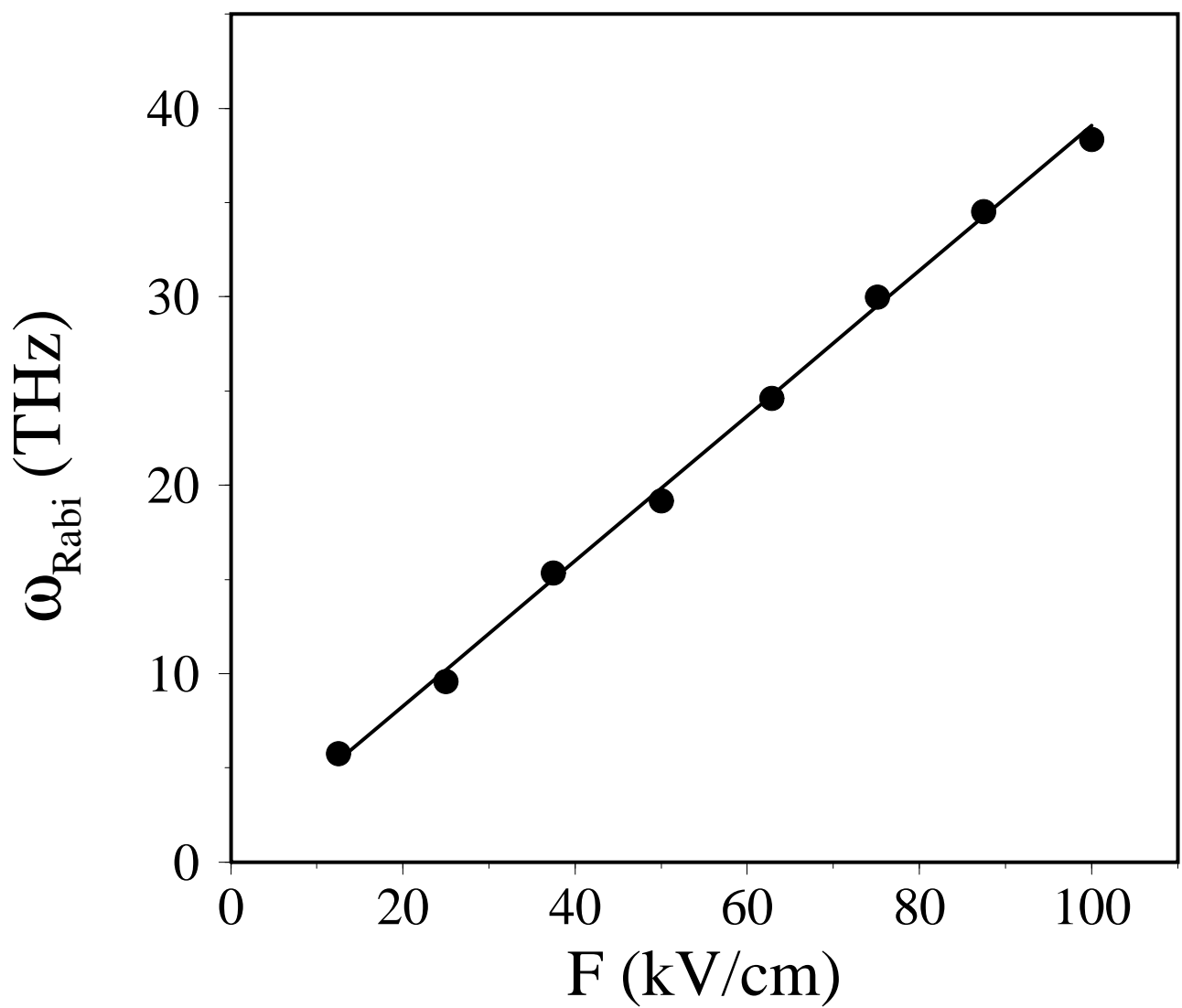
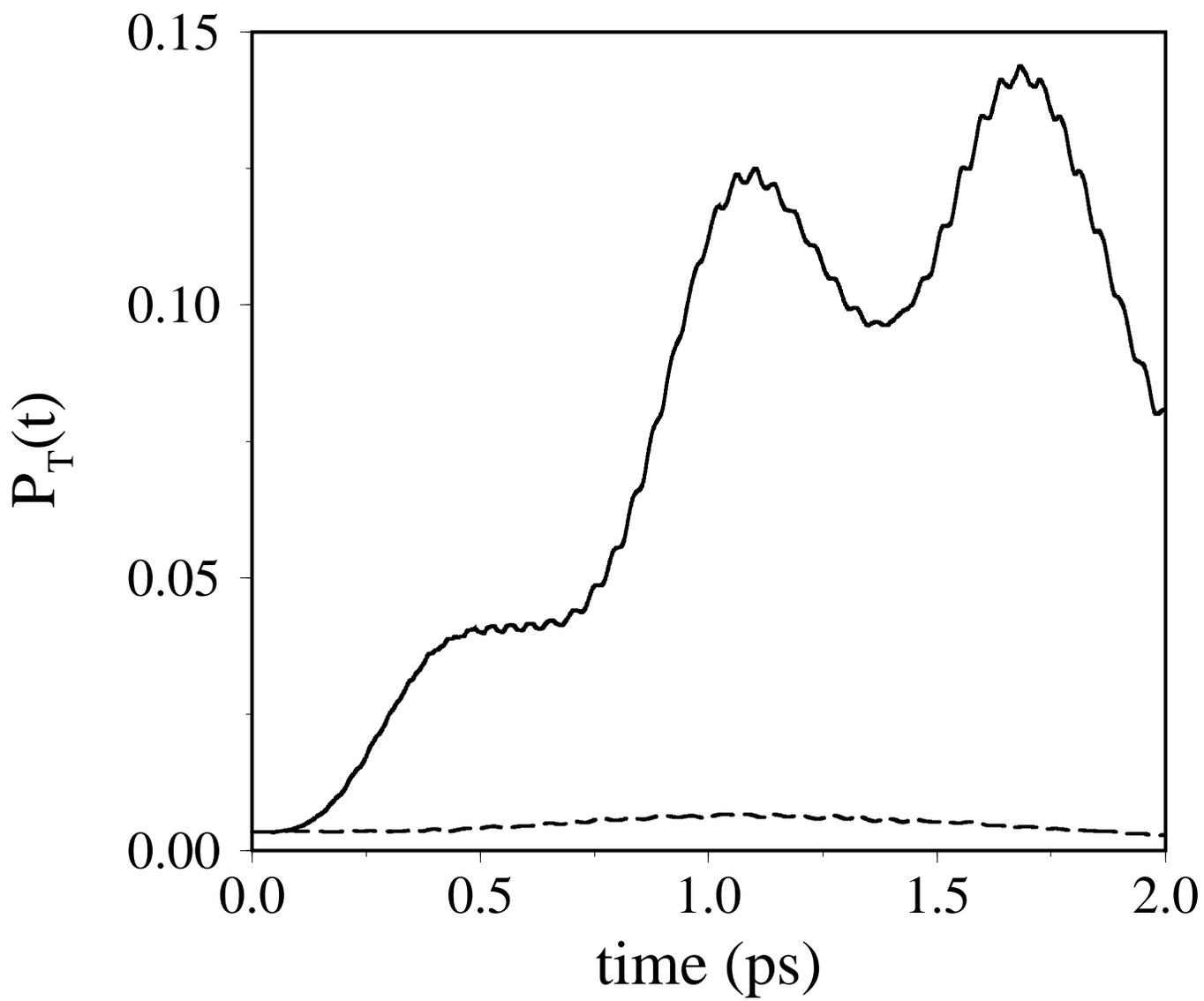
\end{document}